\documentclass[aps, prb, twocolumn, superscriptaddress, amsmath,amssymb]{revtex4}

\usepackage{bm}
\usepackage[dvips]{graphicx}
\usepackage{dcolumn}
\usepackage{longtable}
\newcommand{\NCO}{Na$_{x}$(H$_{3}$O)$_{z}$CoO$_{2} \cdot y$H$_{2}$O}

\begin{document}


\title{Magnetic anomalies of hydrous cobaltate compound \NCO~ detected by NMR and NQR measurements}


\author{Y.~Ihara}
\altaffiliation{ihara@scphys.kyoto-u.ac.jp} 
\author{H.~Takeya}
\affiliation{Department of Physics, Graduate School of Science, Kyoto University, Kyoto 606-8502, Japan.}
\author{K.~Ishida}
\altaffiliation{kishida@scphys.kyoto-u.ac.jp} 
\affiliation{Department of Physics, Graduate School of Science, Kyoto University, Kyoto 606-8502, Japan.}
\author{K.~Yoshimura}
\affiliation{Department of Chemistry, Graduate School of Science, Kyoto University, Kyoto 606-8502, Japan.}

\author{K.~Takada}
\author{T.~Sasaki}
\affiliation{Nanoscale Materials Center, National Institute for Materials Science, 1-1 Namiki, Tsukuba, Ibaraki 305-0044, Japan.}

\author{H.~Sakurai}
\author{E.~Takayama-Muromachi}
\affiliation{Advanced Nano Materials Laboratory, National Institute for Materials Science, 1-1 Namiki, Tsukuba, Ibaraki 305-0044, Japan.}


\date{\today}

\begin{abstract}
In order to investigate the relationship between superconductivity and magnetism in bilayer-hydrate cobaltate \NCO~
Co nuclear magnetic resonance (NMR) and nuclear quadrupole resonance (NQR) measurements were performed on 
three different samples, which demonstrate various ground states at low temperatures. 
The appearance of small internal fields is observed in the NQR spectra below approximately 6 K 
on one of the samples that possesses the largest $c$-axis length and the highest NQR frequency. 
The other two samples exhibit superconducting transition in zero magnetic field, 
while these two samples show different ground states in the magnetic fields greater than 5 T. 
The comparison of the NMR spectra of these two samples obtained in high magnetic fields reveals
the appearance of static internal magnetic fields at the Co site below 4 K in 
the sample that possesses the intermediate $c$-axis length and the NQR frequency. 
\end{abstract}


\maketitle



\section{INTRODUCTION}

Superconductivity was found in the sufficiently water-intercalated cobaltate compound, 
\NCO~ with the bilayer-hydrate (BLH) structure\cite{takada-Nature}. 
It is amazing that superconductivity is observed only in the hydrous phase with the BLH structure. 
Superconductivity has not been observed neither in the monolayer-hydrate nor in the anhydrous compounds. 
The superconductivity is realized on the two-dimensional CoO$_{2}$ layers, which consist of a triangular lattice. 
The superconducting pairing state in the BLH compounds has been considered to be unconventional 
on the basis of the results of the power-law temperature dependence of the specific heat 
\cite{jin-PRB72,oeschler-CJP43,oeschler-PB359,ueland-PC402} 
and the nuclear spin-lattice relaxation rate \cite{ishida-JPSJ72,fujimoto-PRL92}
in the superconducting state. 
Despite the intriguing physical properties both in the superconducting and normal state, 
the details still remain to be solved due to the difficulty in the reproducible sample preparation. 

\begin{figure}
\begin{center}
\includegraphics[width=7cm]{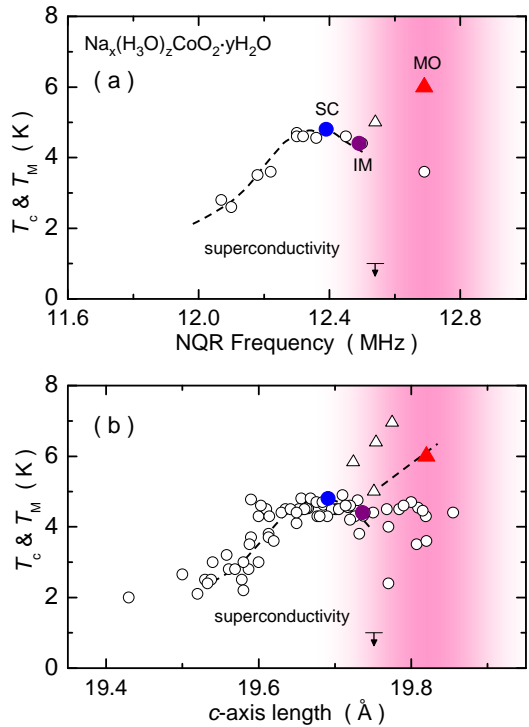}
\caption{
				Superconducting transition temperature $T_{c}$ and magnetic transition temperature $T_{M}$ of various samples 
				reported in the literature 
				\cite{lynn-PRB68, sakurai-JPSJ74, poltavets-PRB74, milne-PRL, jorgensen-PRB68, chen-PRB70, barnes-PRB72, 
				badica-JPCS67, lorenz-PC402, jin-PRL91, jin-PRB72, ohta-JPSJ74, foo-SSC, foo-MRB, schaak-nature424, cao-JPCM15, 
				chou-PRL92, zheng-JPCM, ihara-JPSJ75}.
				These values are plotted against (a) the NQR frequency and (b) the $c$-axis length. 
				The circles and the triangles indicate $T_{c}$ and $T_{M}$, respectively. 
				The down arrow indicates that the superconducting transition was not observed down to 1.5 K. 
				The superconducting transition temperature becomes maximum around $c=19.69$ \AA. 
				The magnetic phase appears in the colored region, where the $c$ axis is approximately $19.8$ \AA. 
				The samples plotted by the filled marks were used in this study. 
				}
\label{Phasediagram}
\end{center}
\end{figure}

In the BLH compounds, the ground state strongly depends on the chemical compositions, 
Na ion ($x$), oxonium ion H$_{3}$O$^{+}$ ($z$) and water molecule ($y$) contents. 
It is difficult to control these parameters precisely during the water intercalation. 
In addition, the water molecules easily evaporate into the air, 
when the samples were preserved in an ambient condition after the water intercalation. 
The unstable nature of the BLH compounds causes the sample dependence of various physical quantities. 
The sample properties have to be clarified in detail both from the microscopic and macroscopic measurements	
before the investigation of the physical properties on the BLH compounds.
In order to summarize the sample dependence of superconducting transition temperatures, 
we have constructed a phase diagram of the BLH compounds 
on the basis of the nuclear quadrupole resonance (NQR) measurements (Fig.~\ref{Phasediagram} (a)). 
In the phase diagram, the Co-NQR frequency $\nu_{Q}$ was used 
as a promising reference for the ground state of the BLH compounds \cite{ihara-JPSJ75-12}, 
because the NQR frequency sensitively reflects the crystalline distortions around the Co site, 
which are induced by the water intercalation. 
Theoretically, the distortions have been predicted to have a relation 
with the formation of superconductivity\cite{yanase-JPSJ74, mochizuki-JPSJ74}.  
This phase diagram has been reproduced by the subsequent experiments \cite{michioka-JPSJ75, kobayashi-JPSJ76}, 
and extended to the higher frequency region. 
It is empirically revealed that the $\nu_{Q}$ detects the sample dependence of parameters 
which are closely related to the formation of superconductivity. 
However, for the full understanding of the relationship between the ground state of the BLH compounds and the NQR frequency, 
the effect of the hole doping to the CoO$_{2}$ layers has to be investigated in detail, 
because the NQR frequency depends also on the concentration of the on-site $3d$ electrons 
in addition to the dominant lattice contributions. 

In this study, we performed the Co-NQR and nuclear-magnetic-resonance (NMR) measurements on three different samples, 
which are located at three typical positions in our phase diagram. 
Some results have been already published in Refs.~[\cite{ihara-JPSJ74}] and [\cite{ihara-PRB75}].
One of the samples, which possesses the longest $c$-axis length and the highest $\nu_{Q}$, 
demonstrates a magnetic transition at approximately 6 K. 
The effects of the small internal fields were observed on the Co-NQR spectra below the temperature\cite{ihara-JPSJ74}. 
The sample that has the shortest $c$-axis length and the lowest $\nu_{Q}$ among the three samples shows the superconducting 
transition at the highest temperature $T_{c} = 4.8$ K. 
In this sample, the nuclear spin-lattice relaxation rate divided by temperature $1/T_{1}T$ was found to 
keep increasing down to 1.5 K without showing any anomalies, 
when the superconductivity is suppressed by the strong magnetic fields \cite{ihara-PRB75}. 
The third sample possesses the intermediate values of $c$-axis length and $\nu_{Q}$. 
In zero fields, the superconducting transition at $T_{c} = 4.4$ K 
and no trace of magnetic anomaly were observed from the NQR measurement on this sample. 
While the physical properties of this sample resemble in those of the superconducting sample in zero fields, 
a weak magnetic anomaly appears on this sample in the magnetic fields greater than 5 T. 
It is considered that the magnetic anomaly appears when the superconductivity is suppressed by the magnetic fields. 
Here, we refer the magnetic ordering sample, the superconducting sample, and the intermediate sample 
as the MO sample, SC sample, and IM sample, respectively. 
The results of the magnetization measurements on the SC and IM samples 
have been already published in Ref.~[\cite{sakurai-PRB74}]. 
The SC and IM samples originate from the same batches as the samples 
referred as NaOH $= 10$ ml and NaOH $= 0$ ml in the report\cite{sakurai-PRB74}, respectively. 
The field-induced magnetic anomaly on the IM sample reminds us of 
the field-induced magnetism in Ce-based heavy Fermion superconductors 
CeRhIn$_{5}$ \cite{knebel-PRB74} and CeCoIn$_{5}$ \cite{young-PRL98}.
It is considered that the superconductivity in cobaltate is also intimately related to the magnetism. 

\section{EXPERIMENTAL}

\subsection{NMR and NQR measurements}

Co-NQR and NMR experiments were carried out with the conventional spin-echo technique. 
The typical values of the pulse widths and the separation between two pulses were 
$\sim 10$ $\mu$s and $30\sim 50$ $\mu$s, respectively.
The NQR spectra were obtained by recording the spin-echo intensity with varying the frequency. 
For the NMR measurements, 
the powder samples were aligned in a strong magnetic field and fixed by an unreactive material, 
which did not degrade the sample quality. 
Due to the anisotropy of susceptibility, crystalline $ab$-plane tends to be parallel to the field direction. 
As a result, magnetic fields were applied within the $ab$ plane. 
The nuclear spin-lattice relaxation time $T_{1}$ was measured with saturation recovery method. 
In order to determine $T_{1}$, the relaxation curves were fitted by the theoretical formula with single $T_{1}$ component. 
The experimental results were sufficiently fitted except for those obtained in the IM sample 
at low temperatures and strong magnetic fields. 
The details are explained in \S~\ref{sec:NQR}.

\subsection{Sample characterization}

It is required to investigate the sample properties carefully and confirm the reproducibility 
on the cobaltate compounds, especially on the hydrous phase. 
In order to compare the physical properties of our samples to those of others, 
the superconducting transition temperatures reported in the literature 
				\cite{lynn-PRB68, sakurai-JPSJ74, poltavets-PRB74, milne-PRL, jorgensen-PRB68, chen-PRB70, barnes-PRB72, 
				badica-JPCS67, lorenz-PC402, jin-PRL91, jin-PRB72, ohta-JPSJ74, foo-SSC, foo-MRB, schaak-nature424, cao-JPCM15, 
				chou-PRL92, zheng-JPCM, ihara-JPSJ75}
are plotted against the $c$-axis length of each sample in Fig.~\ref{Phasediagram} (b). 
A relationship was observed between $T_{c}$ and the $c$-axis length. 
The scattered data points indicate that the $c$-axis length is not the only parameter 
that determines the ground state of the BLH compound. 
The $c$-axis length, however, is considered to be the dominant parameter, 
and can be a useful reference to compare the sample properties of various reports.  
In the figure, superconductivity seems to be suppressed in some samples with $c \sim 19.75$ \AA, 
probably due to the appearance of the magnetism. 
The magnetism was reported to be observed on samples in the colored region \cite{ihara-JPSJ74, sakurai-JPSJ74, higemoto-PB374}. 
We also found some samples, in which both the magnetic and superconducting transitions are observed. 
It has not been revealed yet how these two transitions coexist in one sample. 

\begin{table}[b]
	\centering
	\caption{The chemical and physical parameters of the superconducting (SC) sample, magnetic ordering (MO) sample, and intermediate (IM) sample. }
		\begin{tabular}{l l l l l l}
			\hline
			Sample\; & $T_{c}~(T_{M})$ & $c$ axis (\AA) & Na content & valence & $\nu_{Q}$ (MHz) \\
			\hline 
			SC	& 4.8				& 19.691	& 0.38	& 3.48	& 12.39	 \\
			IM	& 4.4				& 19.736	& 0.33	& 3.48	& 12.49	 \\
			MO	& 3.6 (6.0)	& 19.820	& 0.32	& 3.40	& 12.69	 \\
			\hline
		\end{tabular}
	\label{parameters}
\end{table}%
The powder samples we measured were carefully characterized by using the inductively coupled plasma-atomic emission spectroscopy, 
redox titration, and x-ray diffraction measurements. 
The superconducting transition temperatures were determined by the onset of the Meissner signal 
measured with the {\it in situ} NQR coil ($f \sim 13$ MHz).  
Superconductivity was observed also from the dc magnetization measurements in $10^{-3}$ T with a SQUID magnetometer. 
The superconducting transition temperature $T_ {c}$, Na content $x$, and Co valence $v$ determined by these experiments are 
listed in Table.~\ref{parameters}. 
The oxonium ion content $z$ is estimated from the relationship $z=4-x-v$. 
The $c$-axis parameter, the NQR frequency, and the superconducting transition temperature of the IM sample 
are in-between the SC sample and the MO sample.  
This result suggests that the IM sample would be located in the vicinity of the phase boundary 
between the superconducting and the magnetic ordering phases. 
The filled marks plotted in Fig.~\ref{Phasediagram}~(b) represent the positions for the samples studied here. 

\section{RESULTS}
\subsection{NQR measurements}
\label{sec:NQR}
\begin{figure}
\begin{center}
\includegraphics[width=7.5cm]{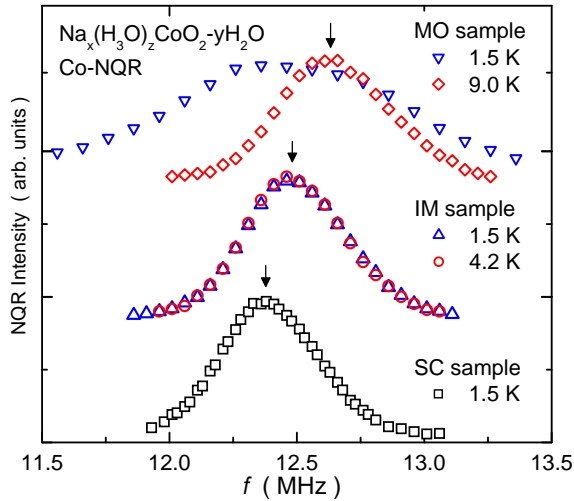}
\caption{	NQR spectra of three samples studied here. 
					The peak positions are pointed by arrows. 
					The spectral broadening was observed only on the MO sample below $6$ K. 
					}
\label{NQRSP}
\end{center}
\end{figure}%

\begin{figure}
\begin{center}
\includegraphics[width=7cm]{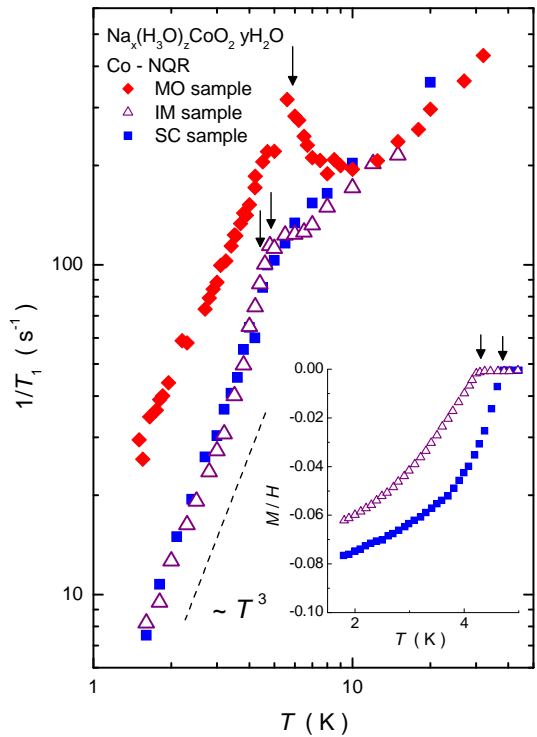}
\caption{	Temperature dependence of $1/T_{1}$ on three samples, which are measured with the Co-NQR technique. 
					The superconducting and magnetic ordering temperatures are pointed by arrows. 
					A diverging behavior was observed only on the MO sample. 
					On the other two samples, power-law behavior was observed below $T_{c}$. 
					The inset shows the temperature dependence of magnetization measured in $10^{-3}$ T with the SQUID magnetometer. 
					}
\label{NQRT1}
\end{center}
\end{figure}

The Co-NQR measurements were performed on three samples, 
and the significant difference was observed in the peak position of the NQR spectra. 
The NQR spectra of three samples are displayed in Fig.~\ref{NQRSP}, 
where the peak positions of each spectrum are pointed by arrows. 
The Co-NQR frequency $\nu_{Q}$ was determined as the resonant frequency of the $m = \pm 7/2\leftrightarrow \pm 5/2$ transitions. 
The values of $\nu_{Q}$ on three samples are shown in Table.~\ref{parameters}. 
On the basis of the NQR frequency obtained here, 
the positions of present samples are pointed by filled marks in our phase diagram (Fig.~\ref{Phasediagram}~(a)), 
where the IM sample was found to be located exactly at the phase boundary between the superconducting and magnetic ordering phases. 
The ground state of the IM sample in zero magnetic field is superconducting, 
because a clear superconducting transition was observed from the magnetization measurements, 
which is displayed in the inset of Fig.~\ref{NQRT1}. 
It should be noted that the Meissner fraction of the IM sample is approximately $80$ \% of that of the SC sample at $1.5$ K. 
Since these two samples were synthesized from the identical starting material, 
the comparable diamagnetic signals indicate that 
the IM and SC samples possess a similar amount of superconducting volume fraction. 
The NQR spectra of the IM sample shown in Fig.~\ref{NQRSP} 
do not depend on the temperature from 4.2 K (circle) to 1.5 K (triangle). 
This result indicates that the static internal fields at the Co site are absent even at $1.5$ K on the IM sample. 
We also noticed that the NQR spectral width of the IM sample is comparable to that of the SC sample (square). 
The sample-independent spectral width suggests that 
the sample inhomogeneity does not cause any difference in the physical properties of three samples. 

The apparent spectral broadening was observed only on the MO sample below $T_{M}\sim 6$ K, 
although the NQR spectrum does not show any clear splitting even at the lowest temperature. 
The spectral broadening is due to the distribution of the weak internal fields at the Co site. 
The distribution was investigated from the NQR spectrum of the MO sample at 1.5 K. 
The details of the analyses are found in a separate paper\cite{ihara-PB403}, 
and explained briefly in \S ~\ref{sec:MO}. 

The magnetic ordering in the MO sample was also confirmed by the other quantity; 
the nuclear spin-lattice relaxation rate $1/T_{1}$, 
which was measured at the peak positions of the NQR spectra. 
The temperature dependence of $1/T_{1}$ is shown in Fig.~\ref{NQRT1}.
The values of $1/T_{1}$ on the MO sample diverge at $T_{M} \sim 6$ K
where the NQR spectra starts to broaden. 
This behavior indicates that the electronic spins slow down toward the ordering temperature, 
because the low-energy excitations of the electronic spins near the magnetic instability 
give rise to the divergence of $1/T_{1}$. 
On the other two samples, the divergence was not observed at all, 
and $1/T_{1}$ starts to decrease just below $T_{c}= 4.4$ K and $4.8$ K for the IM and SC samples, respectively. 
The absence of the Hebel-Slichter peak \cite{coherencepeak} and the power-law temperature dependence at low temperatures  
suggest that the BLH cobaltate is an unconventional superconductor 
with line-nodes on the superconducting gap \cite{ishida-JPSJ72, fujimoto-PRL92}. 
The relaxation curves of the nuclear magnetization were sufficiently fitted by the theoretical function with single $T_{1}$
even in the superconducting state. 
This result confirms that both the SC and IM samples demonstrate superconductivity on the whole part of the samples. 
Concerning with the results of the NQR measurements at zero magnetic field, 
the physical properties of the IM sample resemble in those of the SC sample. 
The difference between the IM and SC samples is observed only in the strong magnetic fields.

\begin{figure}
\begin{center}
\includegraphics[width=7cm]{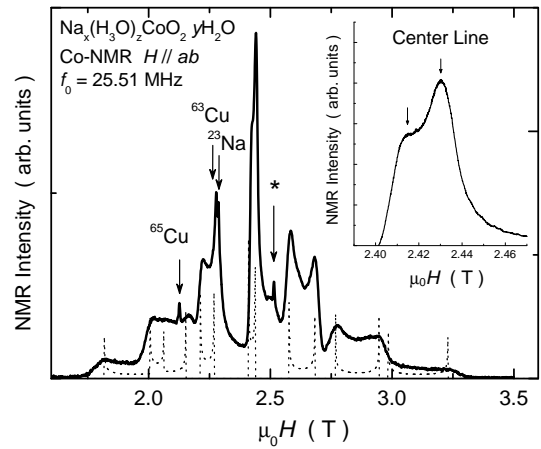}
\caption{	Co-NMR spectrum of the SC sample obtained by applying the external fields in the $ab$-plane. 
					The solid line and dotted line are the experimental result and the simulated result, respectively. 
					The spectral shape is understood as the typical two-dimensional powder pattern. 
					The sharp peaks arising from $^{63}$Cu, $^{65}$Cu in the coil and  $^{23}$Na in the sample are observed 
					at the positions pointed by arrows. 
					The peak marked by asterisk is assigned to be an impurity phase. 
					The inset is the close view of the center line ($m=1/2\leftrightarrow -1/2$ transition). 
					Two-peak structure was observed.
}
\label{NMRSP}
\end{center}
\end{figure}%

\subsection{NMR measurements}
The Co-NMR measurements were carried out in the various external magnetic fields on the IM and SC samples. 
The external fields were randomly applied within the $ab$-plane, 
which is perpendicular to the principal axis of the electric field gradient ($c$ axis). 
The NMR spectrum was consistently explained by a typical two-dimensional powder pattern \cite{pp}. 
The full NMR spectrum of the SC sample is displayed in Fig.~\ref{NMRSP}. 
The dotted line displayed in Fig.~\ref{NMRSP} is the simulated NMR spectrum on the above condition. 
Most of the spectral peaks, except for the sharp peak marked with an asterisk, were accounted for by the calculation. 
We assigned the peak marked with the asterisk as the signal from an impurity phase, 
because the signal has no angle dependence and a long $T_{1}$. 
The candidate for the impurity phase is the paramagnetic metal CoOOH, 
which would be formed by the excess of the water intercalation\cite{sakurai-JPSJ74, ohta-JMMM310}. 
It is noteworthy that no secondary phase was observed in our samples from the x-ray diffraction measurements. 
The signal intensity from the impurity phase was too small to affect the measurements on the center line.
The resonance peaks arising from $^{23}$Na in the sample and from $^{63}$Cu, $^{65}$Cu of an NMR coil were also observed, 
but they were well separated from the center line. 

\begin{figure}
\begin{center}
\includegraphics[width=7cm]{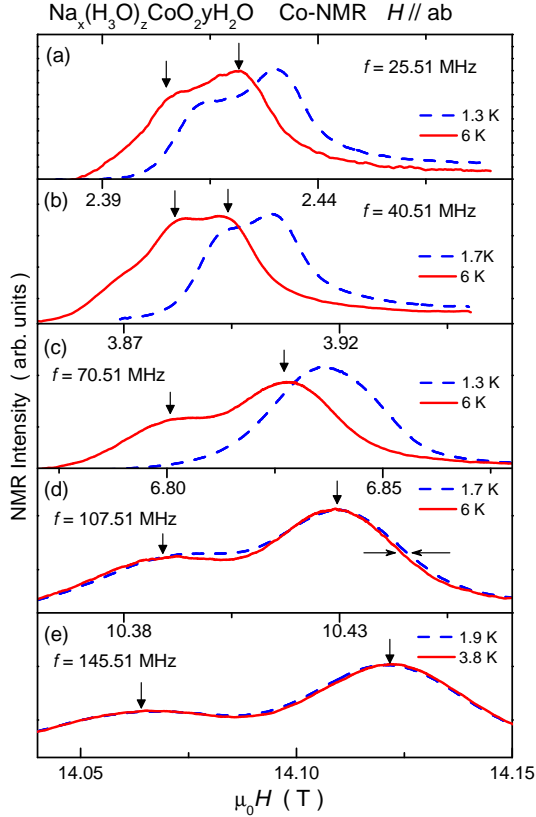}
\caption{	Field dependence of the central peaks on the SC sample.
					(a), (b), (c), (d) and (e) were measured with the frequency of 
					25.51 MHz, 40.51 MHz, 70.51 MHz, 107.51 MHz and 145.51 MHz.
					The solid lines and the dashed lines of each figure are NMR spectra obtained above and below $T_c$, respectively. 
					Peak positions at high temperatures are indicated by down arrows.
					}
\label{SCSP}
\end{center}
\end{figure}

\begin{figure}
\begin{center}
\includegraphics[width=7cm]{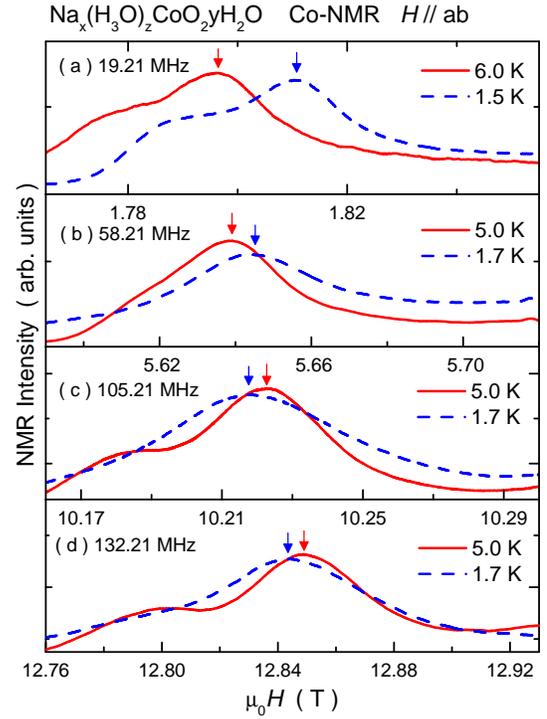}
\caption{	Field dependence of the central peaks on the IM sample. 
					(a), (b), (c) and (d) were measured with the frequency of 
					19.21 MHz, 58.21 MHz, 105.21 MHz and 132.21 MHz, respectively.
					The NMR spectra at the high temperatures ($\sim5$ K) and the low temperatures ($\sim1.7$ K) are shown in each figure. 
					Peak positions of the right peak at high and low temperatures are pointed by down arrows.  
}
\label{IMSP}
\end{center}
\end{figure}

Here, we concentrate on the center line arising from the $m = 1/2 \leftrightarrow -1/2$ transition, 
which is shown in the inset of Fig.~\ref{NMRSP}. 
We found that the center line consists of two peaks pointed by arrows \cite{kato-JPCM18}. 
The double-peak structure is ascribed to the in-plane anisotropy of the electric field gradient and Knight shift. 
The NMR spectra of the SC and IM samples measured at above and below $T_c$ in various fields are shown 
in Figs.~\ref{SCSP} and \ref{IMSP}, respectively. 
The peak positions at $T>T_{c}$ are shown by arrows in Fig.~\ref{SCSP}. 
Within the second order perturbation of the electric quadrupole interaction in the nuclear-spin Hamiltonian, 
the difference of the magnetic fields ($\varDelta H$) 
between the double central peaks at the resonant frequency of $\nu_{0}$ is expressed as
\begin{equation}
\gamma_N \varDelta H = (K_{x} - K_{y}) \nu_{0} + \frac{83}{32}\frac{V_{zz}(V_{xx}-V_{yy})}{(1+K) \nu_{0}}, 
\end{equation}
where $K_{\alpha}$ and $V_{\alpha\alpha}$ ($\alpha=x,y,z$) are Knight shift and electric field gradient 
along $\alpha$ directions, respectively, 
and $\gamma_N$ is the nuclear gyromagnetic ratio. 
The Knight shift is defined as the difference 
between the reference field ($\gamma_{N} H_{0}=\nu_{0} $) and the resonance field divided by the resonance field, 
$K = (H_{0}-H_{\rm res})/H_{\rm res}$.
It should be noted that the resonant frequency ($\nu_{0}$) dependence of the first term, 
which originates from Knight shift anisotropy, 
is opposite to that of the second term which represents the second-order effect of the electric field gradient. 
As increasing fields, 
the difference of the magnetic fields once becomes minimum around 4 T, and it becomes larger again above 7 T. 
The minimum of the splitting width is observed around 4 T, 
because the energy of the quadrupole interaction is comparable to that of the Zeeman interaction around this field. 

The appearance of superconductivity is detected from the shift of the NMR spectra, 
because the in-plane spin susceptibility decreases below $T_{c}$\cite{kobayashi-K, ihara-17O, ihara-JPSJ75, zheng-PRB73}. 
As shown in Figs.~\ref{SCSP} (a), (b) and (c), 
the NMR spectra shift to the higher fields below $T_{c}$ in the fields smaller than 7 T, 
due to the decrease in spin susceptibility in the superconducting state. 
The spectral shift was observed on all the spectral weight. 
This result clearly indicates that the superconductivity is realized in the whole part of the SC sample. 
In 10 T, the shift of a spectral maximum is hardly detectable, 
but a small change in the spectral shape, which is related to the decrease of spin susceptibility, was observed. 
The small change is indicated by the horizontal arrows in Fig.~\ref{SCSP}~(d) at the slope of the spectrum. 
The spectral change becomes smaller than the experimental accuracy in high fields, 
because the change in spin susceptibility becomes subtle near the upper critical field, 
while the spectral width becomes broad in high fields due to the distribution of Knight shift. 
No obvious difference between the NMR spectra at 3.8 K and at 1.9 K in Fig.~\ref{SCSP} (e) shows that 
superconductivity does not occur down to 1.9 K in 14 T.  
It should be noted that the spectral shape does not change at low temperatures and high fields in the SC sample. 
This result indicates that the temperature dependence of Knight shift and electric field gradient is negligibly small 
in this temperature and field ranges. 

In the IM sample, the spectral shift was also observed below $T_{c}$ at a low field ($\sim1.8$ T), 
indicating that the Knight shift decreases in the superconducting state due to the decrease in spin susceptibility. 
It should be noted that the spectral shape and width does not change 
passing through the superconducting transition at this field. 
The invariance of the spectral shape indicates the absence of magnetic transitions in the fields smaller than 2 T.
The shift of the whole spectral weight and the large Meissner fraction, which is comparable to that of the SC sample, confirm that 
the IM sample demonstrates superconductivity on the whole sample at $1.8$ T. 

When the external fields are greater than 5 T, the obvious change in the spectral shape was observed at low temperatures
as shown in Figs.~\ref{IMSP}~(b), (c) and (d). 
The spectral shape varies due to the change in the distribution of the hyperfine fields at the Co site. 
The variation in the spectral shape in the IM sample is apparent at low temperatures, 
when the temperature variation in the spectral shape is compared with those obtained in the SC sample at high fields, 
which are shown in Figs.~\ref{SCSP} (d) and (e). 
The field distribution originating from the vortices is 
estimated to be the order of $10^{-4}$ T at this field range\cite{brandt-PRB68}, 
when the typical penetration depth ($\lambda = 5.68 \times 10^2$ nm) and coherence length ($\xi = 2.32$ nm) 
for the BLH compounds\cite{sakurai-PRB74} are taken into account. 
The spectral broadening in the IM sample is considered to be due to the appearance of static magnetic fields at the Co site. 
The broadening cannot be explained by the magnetism on the small portion of the sample, 
which does not show superconductivity at all. 
In 13 T, the NMR spectral intensity of the IM sample starts to decrease below $4$ K 
due to the short spin-spin relaxation time near the magnetic instability. 
The spectral intensity gradually recovers at low temperatures. 
The smallest spectral intensity is $\sim 30$ \% of that obtained at high temperatures.  
This result suggests that the majority of the sample is in the magnetic state below $4$ K. 
While the peak position shifts to the higher fields below 4 K in 5.6 T,  
the peak position shifts to the lower fields, when $12.8$ T is applied. 
It is considered that the Knight shift increases at low temperatures. 
The details of the temperature dependence of Knight shift are presented in \S \ref{sec:MO} 

\begin{figure}
\begin{center}
\includegraphics[width=7cm]{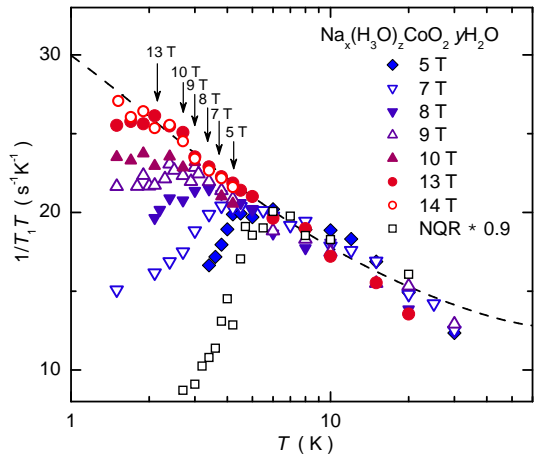}
\caption{
					Temperature dependence of $1/T_{1}T$ in the SC sample at various magnetic fields. 
					The arrows indicate the superconducting transition temperatures at each field, 
					which was determined by the deviation of $1/T_{1}T$ from the normal-state temperature dependence.
					The dashed line is the fitting function described in the text. 
}
\label{SCT1}
\end{center}
\end{figure}%

The temperature dependence of $1/T_{1}T$ was measured on the SC and IM samples in various fields up to 14 T. 
In the SC sample, the relaxation curves were consistently fitted by 
the function for the relaxation of nuclear spins with single $T_{1}$ component. 
This function is written as 
\begin{equation}
\frac{m(\infty)-m(t)}{m(\infty)} = 
A \left( \frac{3}{14}e^{-\frac{3t}{T_{1}}}+\frac{50}{77}e^{-\frac{10t}{T_{1}}}+\frac{3}{22}e^{-\frac{21t}{T_{1}}} \right),
\label{fn:relax}
\end{equation}
where $t$ is the time after the saturation pulse\cite{narath}. 
The coefficient $A$ is a fitting parameter, which possesses the value close to $1$. 
The relaxation-rate measurements were performed on the left-side peak of the center spectrum for the SC sample, 
while $T_{1}$ was measured on the right-side peak for the IM sample. 
The relaxation rate could not be measured on the left-side peak in the IM sample, 
because the double peak structure becomes ambiguous at low temperatures in high fields.  
The similar temperature dependence of $T_{1}$ was observed for both peaks in the SC sample, 
although the absolute values were approximately twice larger on the left-side peak than on the right-side peak. 
The temperature dependence of $1/T_{1}T$ in the SC sample is displayed in Fig.~\ref{SCT1}, 
together with that obtained in zero magnetic field with the NQR measurement. 
The arrows indicate the superconducting transition temperatures in each field, 
which were determined as the temperature where $1/T_{1}T$ starts to deviate from the normal-state temperature dependence. 
When the superconductivity is suppressed by the strong magnetic fields, 
the values of $1/T_{1}T$ continue to increase extending the temperature dependence in the normal state. 
In the SC sample, 
no anomaly other than the superconducting transition was observed from the relaxation measurements even in the field of 14 T. 
This result is consistent with that of the spectrum measurements, 
which does not represent any spectral broadening at low temperatures. 

\begin{figure}
\begin{center}
\includegraphics[width=7cm]{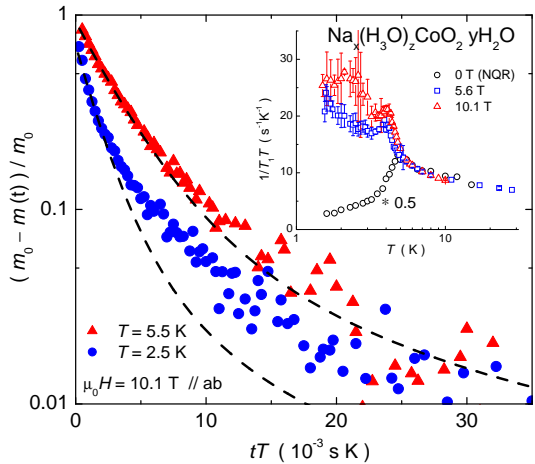}
\caption{	Relaxation curve of the nuclear magnetization on the IM sample.
					The dashed lines represent the theoretical curves for a single $T_{1}$. 
					The experimental data cannot be fitted by the single $T_{1}$ at low temperatures. 
					Inset shows the temperature dependence of $1/T_{1}T$ at 5.6 T, 10.1 T, and 0 T(NQR). 
					The absolute value of $1/T_{1}T$ obtained with NQR measurements was normalized to those with NMR measurement. 
}
\label{IMT1}
\end{center}
\end{figure}%

The physical properties of the IM sample are quite different from those of the SC sample in the magnetic fields. 
At the temperatures above approximately $4$ K, the relaxation curves were consistently explained 
by the theoretical function written in equation (\ref{fn:relax}).  
However, the experimental results could not be fitted by the same function, 
when the IM sample was cooled down below $4$ K. 
The results of the least square fitting to the whole relaxation curves obtained at 2.5 K and 5.5 K are
exhibited in Fig.~\ref{IMT1}. 
The insufficient fit at 2.5 K is considered to be due to the appearance of short-$T_{1}$ components, 
which originate from the ordered magnetic moments. 
In Fig.~\ref{IMT1}, the relaxation curves are plotted against the product of the time and the temperature, 
in order to represent the existence of the short-$T_{1}$ components at low temperatures. 
If a short-$T_{1}$ component contributes to the relaxation curve, the slope would become steep. 
Obviously, the relaxation curve obtained at 2.5 K indicates the presence of short-$T_{1}$ components in the short time regime. 
In addition, the existence of the long-$T_{1}$ components is also observed in the long time regime. 
The relaxation rate in the magnetically ordered state is distributed in space 
probably due to the distribution of the internal fields at the Co site. 
The relaxation curve in the magnetic state can not be explained even with two distinct components (long and short $T_{1}$). 
It is considered that the $T_{1}$ values are continuously distributed. 
The spatial distribution could not be specified with the present NMR experiments on powder samples. 
We fitted all the relaxation curves to the same theoretical function with the single $T_{1}$ component 
even in the magnetically ordered state to determine the typical $T_{1}$ values. 
The least square fits with using full time range are considered to give the relaxation rates on the major fraction of the sample, 
although the results include the large errors below $4$ K. 
The temperature dependence of $1/T_{1}T$ is displayed in the inset of Fig.~\ref{IMT1}. 
The absolute values of the NQR results, which are shown together, were normalized to those of the NMR results. 
The temperature dependence of $1/T_{1}T$ above $T_{c} \sim 4.4$ K is the same among the measurements in three different fields. 

In the normal state, where the relaxation curves were consistently fitted by the theoretical curve, 
$1/T_{1}T$ has a tendency to diverge toward $4$ K, 
and an anomaly is observed at $4$ K, when the superconductivity is suppressed by the strong magnetic fields.  
The temperature dependence in the normal state suggests that a magnetic transition occurs at $4$ K. 
The appearance of the anomaly in the IM sample cannot be explained by the sample inhomogeneity, 
because the anomaly is absent in any fields in the SC sample, 
whose sample inhomogeneity is comparable to that of the IM sample.

\section{DISCUSSION}

\subsection{Quantum Criticality}

In the previous report\cite{ihara-JPSJ75-12}, 
we investigated the temperature dependence of $1/T_1T$ in the normal state from the Co-NQR measurements. 
We found that the sample-dependent behavior of $1/T_{1}T$ can be consistently fitted by 
a function with two fitting parameters $a$ and $\theta$, which is expressed as 
\begin{equation}
 \frac{1}{T_{1}T} = \left( \frac{1}{T_{1}T} \right)_{\rm PG} + \frac{a}{\sqrt{T - \theta}}. 
 \label{fn:SCR}
\end{equation}
The first term $(1/T_{1}T)_{\rm PG}$ is expressed as
\begin{equation}
\left(\frac{1}{T_1T}\right)_{\rm PG} = 8.75 + 15\exp\left(-\frac{250}{T}\right)~~ ( {\rm s}^{-1} {\rm K}^{-1} ),
\end{equation}
which exhibits a pseudo-gap behavior above 100 K.
Here, $a$ is a proportionality constant related to the band structure at the Fermi level and to the hyperfine coupling constant. 
The other parameter $\theta$ is the ordering temperature for the magnetically ordering samples 
and the measure of the closeness to the magnetic instability for the samples without magnetic ordering. 
The pseudo-gap behavior is commonly observed in mono-layered hydrate and non-hydrate cobaltates\cite{ning-PRL93} 
as in BLH compounds above 100 K, 
while the additional second term expressed as $a/\sqrt{T-\theta}$ is observed only in BLH compounds at low temperatures. 

Based on the framework of the self-consistent renormalization (SCR) theory, 
the functional form of the second term is expected 
when the magnetic fluctuations possess three-dimensional antiferromagnetic correlations\cite{moriya-JMMM100}. 
The temperature dependence of $1/T_{1}T$ in the SC sample under the magnetic fields along $ab$ plane  
can be explained by the parameters $a \sim 30$ and $\theta = -0.7 \pm 0.2 $ K, respectively. 
The parameters for the IM sample are $a \sim 10$ and $\theta = 4 \pm 0.03$ K. 
The different $a$ values are due to the angle dependence of the hyperfine coupling constants.

As shown in Fig.~\ref{SCT1}, $1/T_{1}T$ in the SC sample increases with decreasing temperature in the normal state, 
and the increase is interrupted by the onset of superconductivity. 
It is noteworthy that $1/T_{1}T$ continues to increase in the field-induced normal state, 
and follows the fitting curve defined above 5 K. 
This is a great contrast to the Korringa behavior ($T_1T$ = const.), 
which would be observed when the electron-electron interactions are weak enough to form the Fermi-liquid state.  
The continuous increase in $1/T_1T$ down to $T_{c}(H)$ indicates the existence of the strong electronic correlations 
in the SC sample. 
The temperature dependence of $1/T_{1}T$, which is explained by the equation (\ref{fn:SCR}) with $\theta \sim -1$ K, 
strongly suggests that the electronic state of the SC sample is close to the quantum criticality, 
which is characterized by $\theta = 0$ K. 
We expect that the quantum critical point exists in a sample that has the similar chemical composition as the SC sample. 
The precise sample control is required to explore the quantum critical point. 
The temperature dependence of $1/T_{1}T$ reminds us of the specific-heat results on 
CeCu$_2$Si$_2$\cite{gegenwart-PRL81} and CeCoIn$_5$\cite{ikeda-JPSJ70} in magnetic fields, 
both of which are situated at or near the quantum critical point.  

\subsection{Magnetic ordering}\label{sec:MO}

\begin{figure}
\begin{center}
\includegraphics[width=7cm]{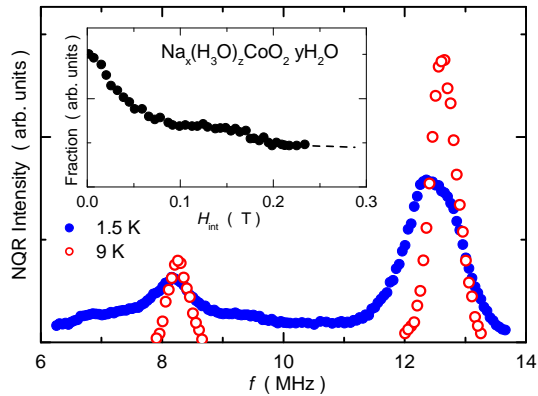}
\caption{	
					The Co-NQR spectra of the MO sample above and below $T_{M} \sim 6$ K. 
					The clear spectral broadening was observed at $1.5$ K.
					In the inset, the estimated internal field distribution is exhibited. 
}
\label{MONQRSP}
\end{center}
\end{figure}%

The broadening was observed in Co-NQR spectra below $6$ K in the MO sample. 
The distribution of the internal fields was estimated from the spectra. 

In general, the nuclear spin Hamiltonian is written as 
\begin{equation}
\mathcal{H}	= \frac{e^2qQ}{4I(2I-1)}\Big\{ \left( 3I_{z}^2-{\bm I}^2\right) + \eta \left( I_{+}^2+I_{-}^2\right) \Big\} + \gamma\hbar{\bm I}\cdot{\bm H}_{\rm int}, 
\label{fn:MO}
\end{equation}
where $eq$ and $Q$ are the electric field gradient and the nuclear quadrupole moment, respectively. 
The first term of the right hand side expresses the electric quadrupole interaction, 
and the second term indicates the Zeeman interaction originating from the internal fields. 
Below a magnetic ordering temperature, 
the Zeeman interaction emerges and increases due to the appearance of the internal fields at the nuclear site. 
The relationship between the resonant frequency and the internal fields for the MO sample is derived from this Hamiltonian. 

First, we determined the direction of the internal field to $H_{\rm int} \perp c$, 
where the $c$ axis is the principal axis of the electric field gradient, 
because the resonance peak arising from $m = \pm 5/2 \leftrightarrow \pm 3/2$ transitions become broader than that
arising from $m = \pm 7/2 \leftrightarrow \pm 5/2$ transitions. 
If the internal fields were along the $c$ axis, 
$\pm 7/2 \leftrightarrow \pm 5/2$ transition line would become the broadest within the three NQR lines. 
Obviously, this is not a case. 
The anisotropic broadening of $\pm 7/2 \leftrightarrow \pm 5/2$ transitions below $6$ K can also be  
consistently explained by assuming $H_{\rm int} \perp c$. 
These two results suggest that the internal fields direct to the $ab$ plane. 

Next, the intensity of the internal fields is estimated using the relationship 
between the resonant frequency and the internal field strength, which is derived from the equation (\ref{fn:MO}). 
The details of the analyses are described in the separate paper\cite{ihara-PB403}. 
For the estimation of the fraction, we used the NQR spectra in the frequency range of $7 \sim 10.5$ MHz, 
because almost linear relationship was observed between the frequency and the internal field strength in this frequency range. 
The result is exhibited in the inset of Fig.~\ref{MONQRSP}. 

It was found that the maximum fraction is at zero field, and that rather large fraction is in the small field region. 
We also point out that a weak hump was observed on the fraction around $0.15$ T, 
which corresponds to 0.1 $\mu_{\rm B}$ when we adopt 1.47 T$/\mu_{\rm B}$ as the coupling constant \cite{kato-JPCM18}. 
It should be noted that NQR measurements were performed on the Co site, where the magnetic moments are located. 
The distribution of the hyperfine fields at the Co site suggests that 
the size of the ordered moments has the spatial distribution. 
The magnetism in the BLH compounds is considered to be a spin-density-wave type 
with the modulation on the size of the ordered moments. 
In order to investigate the magnetic structure in detail, neutron diffraction measurements are required.

\begin{figure}
\begin{center}
\includegraphics[width=7cm]{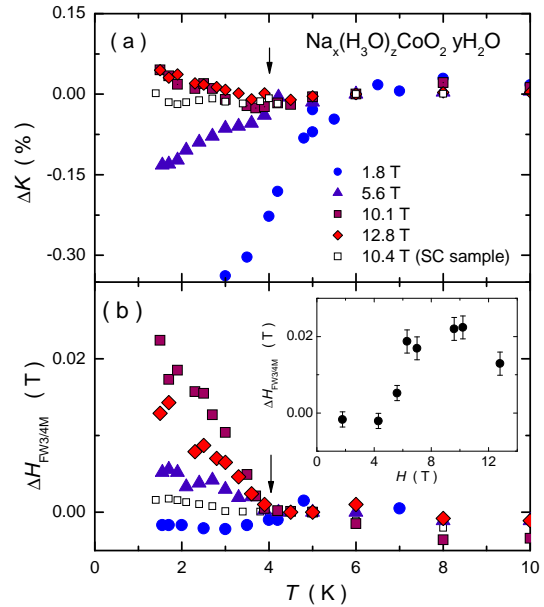}
\caption{
					(a) Temperature dependence of Knight shift measured in four different fields on the IM sample and $10.4$ T on the SC sample. 
					(b) Temperature dependence of the spectral width, which was characterized by the full width at three quarters maximum (FW3/4M). 
					The symbols are the same as those in (a). 
					A magnetic anomaly was observed at $4$ K in the IM sample. 
					(b)-inset. Field dependence of FW3/4M at 1.5 K on the IM sample.
}
\label{shift}
\end{center}
\end{figure}%

The internal fields are observed also in the IM sample,
when the magnetic fields greater than $5$ T are applied. 
The Knight shift of the IM sample scaled at $6$ K 
\begin{equation}
 \varDelta K(T)=K(T)-K(6\;{\rm K})
\end{equation}
is displayed in Fig.~\ref{shift} (a), 
where $\varDelta K(T)$ measured on the SC sample at $10.4$ T is shown together. 
Compared with the result of the SC sample, the Knight shift on the IM sample slightly increases below $4$ K. 
It seems that the magnetic structure has a ferromagnetic component, 
because the magnetization increases below $4$ K. 
The temperature dependence of the magnetization measured on the IM sample was presented in Ref.~[\cite{sakurai-PRB74}]. 
The increase in the magnetization was also observed below approximately $4$ K, 
when the external fields greater than $5$ T are applied. 
The resemblance of the temperature dependence between the bulk magnetization and the Knight shift suggests that 
the increase in Knight shift is an intrinsic behavior.

In addition to the temperature dependence of $\varDelta K$, 
we investigated the temperature dependence of the spectral broadening in several fields 
by evaluating the spectral width at the three-quarter intensity of the highest peak.  
We used the full width at the three-quarter maximum (FW3/4M) in order to 
minimize the spectral width arising from the in-plane anisotropy of the Knight shift and the electric-field gradient. 
These contributions are small near the peak intensity. 
The line width estimated near the peak intensity originates mainly from the distribution of the hyperfine fields at the Co site. 

We defined the value $\varDelta H_{\rm FW3/4M}$ as the difference of FW3/4M from that at $5$ K 
\begin{equation}
 \varDelta H_{\rm FW3/4M} (T) = H_{\rm FW3/4M}(T) -H_{\rm FW3/4M}({\rm 5K}), 
\end{equation}
which are plotted as a function of temperature in Fig.~\ref{shift} (b).  
The spectral broadening due to the electric field gradient is completely canceled by subtracting the FW3/4M at $5$ K, 
because the electric field gradient does not depend on temperatures below $50$ K. 
The temperature dependence of $\varDelta H_{\rm FW3/4M}$ possesses the magnetic origin. 
At $1.8$ T, 
the spectral width slightly decreases below $T_{c}$, because the spin susceptibility decreases in the superconducting state. 
The decrease in the spectral width in the superconducting state is observed 
when the spectral width is determined mainly by the inhomogeneity of the Knight shift. 
This result indicates that the magnetic-field distribution generated by the vortices in the superconducting state 
is smaller than those arising from the distribution of Knight shift even in $1.8$ T. 
In the fields greater than $5$ T, the FW3/4M starts to increase below $4$ K, 
where the anomaly was also observed in the temperature dependence of $1/T_{1}T$. 
It is considered that the static internal fields cause the broadening on the NMR spectra of the IM sample. 
The field dependence of the onset temperature, 
where the spectra start to broaden, is considerably weak in the field range $5-14$ T. 
The FW3/4M of the SC sample measured in $10.4$ T does not represent any anomaly at $4$ K, 
but shows a gradual increase, which is roughly proportional to the bulk susceptibility.

The spectral width at $1.5$ K $\varDelta H_{\rm FW3/4M}(1.5 ~\rm{K})$ is plotted as a function of external fields 
in the inset of Fig.~\ref{shift} (b). 
A clear increase is observed at 4 T, and the values are saturated above 6 T. 
The saturated value is approximately 0.02 T, which corresponds to the ordered moment of 0.01 $\mu_{\rm B}$ 
by assuming the hyperfine coupling constant being 1.47 T/$\mu_{B}$\cite{kato-JPCM18}. 
It can be considered that the weak magnetic ordering, which possesses ordering temperature $T_{M} \sim 4$ K, is masked by 
the superconductivity with $T_{c} \sim 4.4$ K. 
The magnetism appears when the superconductivity is suppressed below $T_{M}$ by external magnetic fields. 
The emergence of the internal fields in the external magnetic fields and the clear superconducting transition in zero field 
indicate the intimate relationship between superconductivity and magnetism. 
The similar field-induced magnetism has been observed in cuprate and heavy-Fermion superconductors, 
in which superconductivity occurs near the magnetic instability \cite{knebel-PRB74,lake-Nature415}.

Intriguingly, the spectral broadening was observed at the same time as 
the decrease in the Knight shift due to superconductivity at 5.6 T. 
In this field, superconductivity coexists with magnetism within the IM sample.  
One might consider that the IM sample includes two different phases, 
which are the superconducting phase and the magnetic phase. 
This possibility, however, can be ruled out by the large superconducting volume fraction of the IM sample at low fields, 
which is comparable to the SC sample, 
and the obvious change in the NMR spectra at high fields. 
The IM sample varies the ground state depending on the external fields. 
At the intermediate fields $\sim5$ T, superconductivity coexists with magnetism microscopically. 
There are two candidates to comprehend the relationship between magnetism and superconductivity in the IM sample. 
One is the uniform coexistence, where the ordered moments appear in the superconducting state. 
The other possibility is that the magnetism occurs in the region, 
where the superconductivity is suppressed, such as the vortex cores, and is apart from the superconducting region. 
In the present study, the spectral broadening with magnetic origin was not observed in the superconducting state at low fields, 
and was suddenly observed when the superconductivity was suppressed below approximately $4$ K. 
Our experimental result might suggest that the magnetism cannot coexist with superconductivity uniformly, 
but occurs in the vortex core. 
However, it is difficult to observe spatial distribution of the electronic state from the broad NMR spectra of the powder samples. 
NMR measurements on single crystals are required to obtain the sharp NMR spectra and 
to investigate the spatial distributions of $T_{1}$ in high fields.

\section{CONCLUSION}
We performed Co-NQR and NMR measurements on three BLH \NCO~ samples, which demonstrate various ground states. 
The experimental results presented here suggest that the superconductivity in BLH compound is intimately related to the magnetism. 
Since a subtle change in the chemical compositions strongly affects the physical properties, 
it is required to characterize the samples from both macroscopic and microscopic measurements. 
A magnetic transition was observed in zero magnetic field in the MO sample. 
In the MO sample, it was found that the spatially distributed internal fields appear below $6$ K. 
In the SC sample, which shows the highest $T_{c} = 4.8$ K, 
the magnetic fluctuations were found to keep increasing down to $1.5$ K, 
when the superconductivity was suppressed by the strong magnetic fields. 
The temperature dependence of $1/T_{1}T$ suggests that the SC sample is located in the vicinity of the quantum critical point. 
In the IM sample, which possesses the intermediate parameters, 
a magnetic anomaly was observed only in the magnetic fields greater than $5$ T, 
while this sample shows the clear superconducting transition at $4.4$ K and no trace of the magnetism at zero field. 
We point out that the physical properties of the IM sample in the magnetic fields are quite different 
from those of the SC sample, 
although the properties in zero magnetic field are similar. 
It seems that the magnetism in the IM sample appears in the field-induced normal state, such as at the vortex core, 
but our experimental resolution on the powder sample was not enough to determine how these two phases coexist. 
Co-NMR measurements using single-crystalline sample are needed to explore the details of this magnetic anomaly. 

\begin{enumerate}
	\item 
\end{enumerate}

\begin{acknowledgments}
We thank H.~Ohta, C.~Michioka, Y.~Itoh, H.~Yaguchi and Y.~Maeno for experimental support and valuable discussions. 
We also thank H.~Ikeda, K.~Yamada, Y.~Yanase, M.~Mochizuki and M.~Ogata for valuable discussions.
This work was partially supported by CREST of the Japan Science and Technology Agency (JST) and 
the Grant-in-Aid for the Global COE program ``The Next Generation of Physics, Spun from University and Emergence'' from 
the Ministry of Education, Culture, Sports, Science and Technology (MEXT) of Japan, 
and by Grants-in-Aid for Scientific Research from the Japan Society for the Promotion of Science (JSPS)(No.16340111 and 18340102) and MEXT(No.16076209).
\end{acknowledgments}

\end{document}